%% file: main_new.tex
\documentclass[journal]{IEEEtran}
\ifCLASSINFOpdf \else \fi

\usepackage{graphicx}
\usepackage[cmex10]{amsmath}
\usepackage{subcaption}
\usepackage{amssymb}

\usepackage[table,xcdraw]{xcolor}
\usepackage{multirow}
\usepackage{acro}
\input{acro.tex}
\usepackage{algorithmic}
\usepackage{makecell}
\usepackage{ragged2e}
\usepackage{array}
\usepackage{tabularx}
\usepackage{multirow}
\usepackage{array}

\usepackage{url}
\usepackage{cite}
\usepackage[compact]{titlesec}
\titlespacing{\section}{1.0pt}{*1.0}{*0}
\titlespacing{\subsection}{1.1pt}{*1.1}{*0}
\titlespacing{\subsubsection}{0.3pt}{*0}{*0}
\begin{document}
\bstctlcite{IEEEexample:BSTcontrol}
\title{Sparsity and Resolvability: Re-evaluating Channel Representations For Next Generation Networks}

\author{
\IEEEauthorblockN{Hamza Haif,~\IEEEmembership{Graduate~Student~Member,~IEEE}, Abdelali Arous,~\IEEEmembership{Graduate~Student~Member,~IEEE}, and H\"{u}seyin Arslan,}
\IEEEmembership{Fellow, IEEE}
\thanks{The authors H. Haif,  A. Arous, and H. Arslan, are with the Department of Electrical and Electronics Engineering, Istanbul Medipol University, Istanbul, 34810, Turkey, (e-mail: hamza.haif@std.medipol.edu.tr; abdelali.arous@std.medipol.edu.tr;  huseyinarslan@medipol.edu.tr). }
}

\maketitle

% \begin{abstract}
% % \textcolor{red}{As wireless networks transition toward 6G, the propagation and hardware regimes increasingly challenge classical assumptions on channel sparsity, resolvability, and stationarity due to high-mobility scenarios, clustered scattering, and hardware-induced impairments. As a result, performance assessments based solely on visual sparsity or nominal delay–Doppler separation risk masking fundamental limitations introduced by receiver processing, sampling granularity, and leakage mechanisms. This article provides a comprehensive, signal processing centric framework to deconstruct the coupled nature of sparsity, resolvability, and selectivity across diverse multiplexing domains. We shift the perspective from intrinsic channel physics to observable properties shaped by energy concentration, inter-component correlation, and processing-induced leakage. By evaluating these dynamics within an Extended Vehicular A (EVA) channel profile, we demonstrate how domain-specific representations impact the performance of communication, sensing, and physical layer security. Finally, we propose a paradigm shift toward adaptive interchanged-domain radios, where the processing domain is dynamically changed based on real-time indicators to ensure robustness in non-stationary 6G environments.}
 \begin{abstract}
As wireless networks transition toward 6G, high mobility, clustered scattering, and hardware impairments increasingly challenge classical assumptions on channel sparsity, resolvability, and stationarity. In these regimes, performance assessments based on apparent sparsity or nominal delay and Doppler separation can be misleading, since finite observation, sampling granularity, windowing, and fractional delay or Doppler spreading introduce coupling and leakage that reshape the effective channel seen by the receiver. This article provides a signal processing centric framework that links sparsity, resolvability, and selectivity through receiver observable indicators, including the fraction of power captured by dominant coefficients, the level of coefficient correlation, the effective delay and Doppler resolving capability over the observation window, and processing induced leakage. Building on these observations, we propose an interchanged domain frame concept principle, where the representation and the degree of component separation are adapted according to the propagation regime, the effective SNR under impairments, and the application objective. Using the Extended Vehicular A channel profile as a running case study, we show how different representations lead to different equalization and detection behavior, with implications for communication, sensing, and physical layer security.
\end{abstract}

\IEEEpeerreviewmaketitle
\vspace{-1mm}
\section{Introduction}

\IEEEPARstart{H}{istorically}, wireless system design has relied on time- and frequency-domain representations which maps the signal directly to the physical resources. Particularly, \ac{OFDM}, a fundamental waveform for the fourth- and fifth-generation cellular networks, enables efficient resource allocation under quasi-stationary channel conditions. While this framework has proven effective in terms of scheduling, \ac{CE}, equalization and \ac{MIMO} compatibility, it
fails to adhere to the \ac{6G} scenarios like \ac{UAV}, \ac{NTN}, and high mobility communication. Recent research has explored alternative logical domains such as \ac{DD} and affine representations to better align channel observation with the underlying physics of propagation. \ac{DD}-based waveforms, including \ac{OTFS}, seek to exploit the sparsity of the channel spreading function by localizing \ac{MPCs} along delay and Doppler axes in a 2-dimensional grid \cite{10891132}. Alternatively, chirp-based waveforms, such as \ac{OCDM} and \ac{AFDM}, couple time and frequency through controlled spreading in a 1-dimensional plane, offering robustness against Doppler and enabling new trade-offs between diversity, leakage, and overhead \cite{10769778}. While these approaches demonstrate clear advantages in selected scenarios, their performance critically depends on whether \ac{MPCs} sparsity and resolvability are preserved within their corresponding multiplexing domain. Furthermore, despite these features, signal projections in these domains remain contingent on the assumptions of sparsity and the resolvability of uncorrelated \ac{MPCs}, while the role of sampling granularity, windowing, guards, and leakage on the observable channel selectivity has received less attention, \textcolor{black}{with selectivity referring to \ac{MPCs} magnitude and phase variation within the multiplexing domain.} In 6G propagation scenarios, however, these assumptions are increasingly violated. High mobility, clustered scattering, fractional delay and Doppler effects, hardware impairments, pulse shaping, and guard constraints fundamentally alter the structure of the effective channel \cite{9799524}. Therefore, channel sparsity, \ac{MPCs} resolvability, and multiplexing domain phase selectivity conditions map beyond the intrinsic channel properties to the pulse shaping leakage, hardware impairments, channel modeling accuracy and domain-specific representations. The exact conditions under which these assumptions hold remain insufficiently specified, and their relevance across other multiplexing domains has yet to be thoroughly established. While time and frequency domains representations of sparsity and \ac{MPCs} resolvability have been extensively examined \cite{liu2024channel}, there is a notable gap concerning their manifestation, and implications in emerging multiplexing domains. A key challenge, therefore, lies in the absence of a unified, clear and signal processing centric framework for assessing when and why a given channel representation is sparse, resolvable, or static. Threshold based sparsity decided by coefficients exceeding a noise relative threshold can be misleading when \ac{MPCs} are correlated or when fractional induced leakage persists. Likewise, delay or Doppler separation does not guarantee resolvability if components are strongly correlated or smeared by windowing and hardware impairments \cite{11269017}. These ambiguities complicate performance evaluation and limit the generality of conclusions drawn from idealized models. Although the gini-index, diversity levels and the Ricean K factor have proven to be reliable parameters to evaluate channel sparsity, the correlation between \ac{MPCs} and inter-components leakage are often overlooked \cite{9373010}. In this context, the following questions must be addressed:
% %%%%%%%%%%%%%%%%%%%
\begin{figure*}[t]
    \centering
    \includegraphics[width=1\textwidth,height=6.0cm]{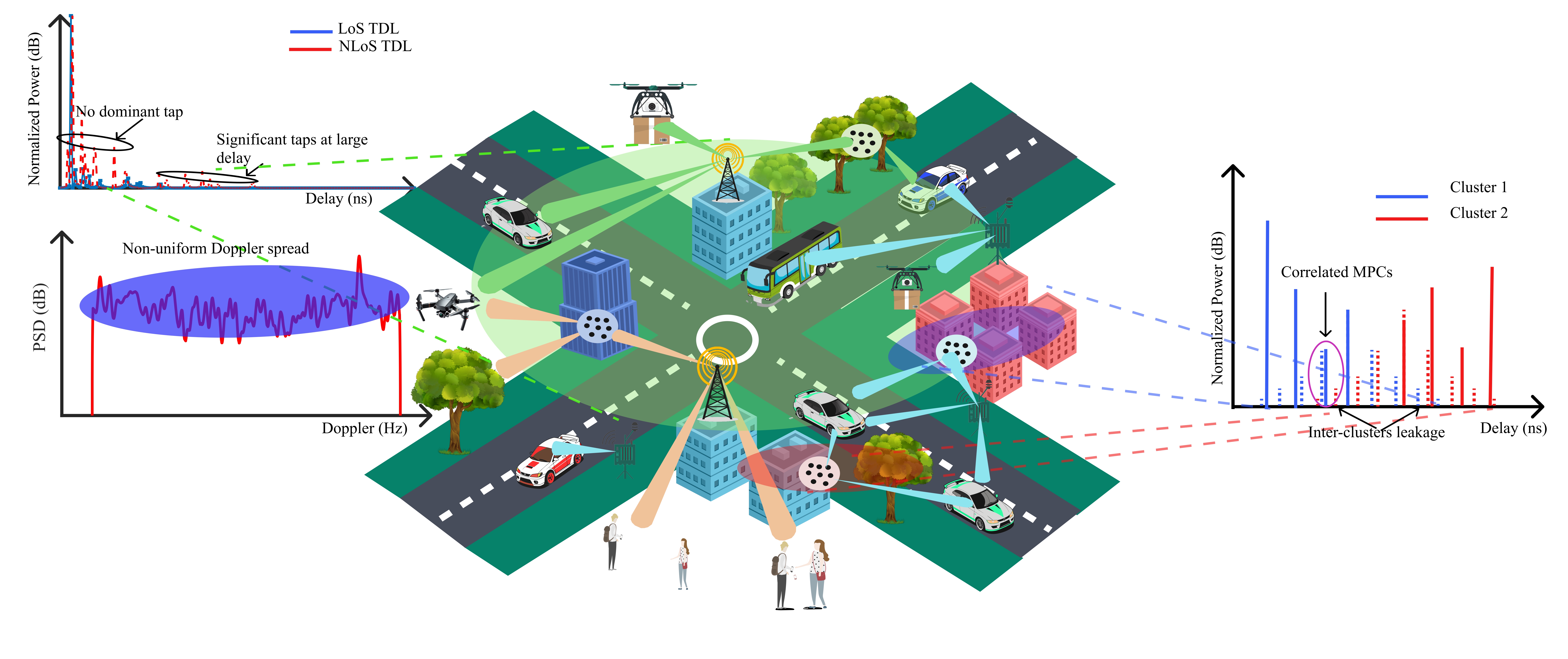}
    \caption{6G vehicular propagation illustration highlighting channel effects relevant to domain assessment, including multipath delay clustering, Doppler spread, and LoS/NLoS profiles.}
    \label{fig:channel_system}
\end{figure*}
% %%%%%%%%%%%%%%%%%%%%
\begin{itemize}
    \item When does threshold based sparsity in a given multiplexing or processing domain translate into reliable equalization and detection gains, and when is it rendered misleading by correlated multipath components and fractional delay or Doppler leakage under practical signal processing?
    \item What governs channel resolvability beyond nominal resolution limits, including discretization, pulse shaping, and hardware impairments, and how do these factors reshape energy dispersion and \ac{MPCs} correlation across the time, frequency, delay-Doppler, and affine domains?
    \item How can these observations be distilled into a low-overhead, domain-centric signal processing selection rule,  targeting communication, sensing, and \ac{PLS} use cases?
\end{itemize} 
\par This magazine presents a structured answer to these questions by providing a signal processing based explanation of sparsity and resolvability, with emphasis on their interrelated effects. These challenges are addressed by rethinking channel sparsity, resolvability, and spreading as coupled and observable properties shaped jointly by propagation and signal processing. Rather than treating them as domain-invariant, we adopt signal processing indicators such as the fraction of power captured by dominant coefficients, the degree of the coefficients correlation, the effective delay and Doppler resolving capability over the signal bandwidth and duration, and the leakage power induced by pulse shaping. Using an \ac{EVA}-like mid-band channel as a representative study case, we demonstrate how these factors influence communication reliability, sensing performance, and physical-layer security in different domains. Following that, we propose an adaptive interchanged-domain radio, where the processing domain is selected dynamically based on pilot-assisted indicators rather than fixed waveform assumptions. 
%The remainder of this article is organized as follows. Section II establishes formal, receiver-centric definitions of sparsity and resolvability and clarifies their intrinsic interdependence. Section III examines how these properties evolve when the channel is projected into conventional and emerging domains. Section IV interprets the implications for communication, sensing, and security, highlighting domain-dependent trade-offs. Section V discusses open challenges related to high-frequency operation, hardware impairments, and AI-enabled receivers. Conclusions are drawn in Section VI.

\section{Sparsity and Resolvability Interplay}
This section investigates the fundamental conditions under which propagation channels exhibit sparsity and resolvability. Without loss of generality, we formalize the conditions under which the \ac{MPCs} representation may be regarded as jointly sparse and resolvable, and clarify the mechanisms governing the coupling between these two properties.
\par \textbf{Channel sparsity}: defines the energy concentration within a chosen representation, rather than a simplistic count of the number of non-zero coefficients or the presence of diagonal structures in an effective channel matrix \cite{liu2024channel}. Such structural indicators are known to be fragile under mobility and Doppler-induced dispersion. A propagation channel that appears underspread and hence sparse in the time domain may simultaneously exhibit significant spreading and irregularity in the frequency domain. This effect is particularly pronounced at sub-6 GHz, where Doppler often manifests as a spectrum of frequency shifts rather than a single deterministic shift. A meaningful sparsity characterization must therefore quantify how the channel's energy is distributed among the dominant components while assessing the degree of statistical dependence between them. In practice, channels dominated by a strong \ac{LoS} component accompanied by weak, diffuse echoes can remain effectively sparse, even when multiple coefficients are non-zero, since the majority of the energy is concentrated in a small number of uncorrelated components. Conversely, a \ac{NLoS} propagation composed of a few dominant yet strongly correlated scattering clusters may exhibit low effective sparsity despite a reduced apparent tap count, as the channel energy is distributed across several mutually dependent components, as depicted in Fig. \ref{fig:channel_system}. Consequently, sparsity metrics based on bounded energy concentration per tap provide a more robust and physically meaningful description than raw coefficient counts, since they implicitly capture the interplay between spreading, relative power, and correlation.
%%%%%%%%%%%%%%%%%%%%
\par \textbf{Channel resolvability}: is a combination of the physical channel, signal properties, and receiver processing that determines whether two \ac{MPCs} can be treated as distinct without incurring significant mutual leakage. In this context, nominal separation in delay or Doppler alone is insufficient to guaranty resolvability. Instead, resolvability is governed by the combined effect of the following key factors:
\begin{enumerate}
    \item \textit{Signal's granularity}: the resolution of the transmitted signal dictates the minimum distinguishable spacing between MPCs. Delay resolution is primarily determined by the signal bandwidth, whereas Doppler resolution is defined by the observation interval over which the channel is probed. 
    \item \textit{Processing-induced leakage}: practical transmitter and receiver operations such as windowing, pulse shaping, and the presence of fractional delay and Doppler components introduce energy spreading that smears MPCs across adjacent resolution bins. This leakage degrades separability even under ideal spacing conditions.
    \item \textit{Mutual correlation}: statistical dependence between MPCs, arising from intra-cluster and inter-cluster leakage can fundamentally limit resolvability despite apparently adequate nominal spacing.
\end{enumerate}
% %%%%%%%%%%%%%%%%%%%
%  \begin{figure}[t]
%      \centering
%      \includegraphics[scale=0.38]{figures/channel_mapping.jpg}
%      \caption{Channel projection from time-frequency to the emerging domains like the delay-Doppler domain and the affine domain.}
%      \label{fig: channel_projec}
% \end{figure}
% % %%%%%%%%%%%%%%%%%%%%
A practical ambiguity follows in the frequency domain, where multiple resolvable taps may originate either from a single physical reflector observed under multiple angles of arrival or from multiple distinct reflectors associated with different Doppler shifts. Only the latter case contributes independent channel information and diversity. Consequently, resolvability assessments based solely on delay or Doppler resolution are incomplete, and the correlation structure among MPCs must be explicitly considered to accurately characterize the channel’s effective resolvability.
\par Sparsity and resolvability are intrinsically interconnected and should be interpreted through the merge of physical channel and signal processing rather than treated as independent, domain-invariant attributes. Channels that appear sparse under coefficient-count and threshold based metrics may nonetheless be difficult to resolve when dominant components exhibit strong mutual correlation or when processing-induced leakage is significant. Therefore, the common assumption of \textit{“sparse and fully resolvable”} typically holds only when a small number of \ac{MPCs} arrive from well-separated clusters, each behaving approximately as a single ray with a smooth power decay, such that the intra-cluster overlap is limited and mutual correlation remains low.

\section{Projection in Different Domains}
This section examines channel features across different representative domains, with system compatibility and resource constraints considerations. 
\subsection{Conventional Domains}
Up to 5G, waveform and channel representation were centered on time and frequency domains, which map directly to the signal lattice. This interaction enabled flexible resource allocation, interference management, and efficient scheduling for OFDM and DFT-s-OFDM.
\par \textbf{Time Domain:} Temporal characterization of the multipath propagation delays encompasses both the signal’s propagation time to the receiver and the number of signal replicas originating from the channel. The maximum excess delay refers to the interval between the first and last detectable MPCs for a given receiver sensitivity. Additionally, tracking the phase of a given MPC over short intervals offers a sense of local stationarity, yet the observation itself does not reveal whether phase alterations arise from Doppler or from the birth and death phenomena in which physical rays appear and vanish within a cluster in dynamic conditions \cite{birth_death}. Moreover, the energy spreading and windowing tails cause fractional delays that leak into adjacent \ac{MPCs} and clusters. Even when \ac{MPCs} appear separated, elevated correlation and leakage reduce practical resolvability.
\par \textbf{Frequency Domain:} The frequency response encapsulates variation in the complex amplitude across the occupied bandwidth by converting the time-domain MPCs to an overlapping spectral view, thereby reflecting varying frequency selectivity levels. This frequency selectivity becomes more severe with larger excess delay and tends to be irregular when MPCs are unevenly spaced in time-domain. Furthermore, distinct Doppler spectra can be observed across different scenarios:

\begin{enumerate}
    \item A single and powerful MPC formed by overlapping rays arriving omni-directionally, produces a Doppler spectrum whose width depends on angular spread ~\cite{iqbal2009generalised}. Nonetheless, with directional antennas at higher carrier frequencies, a single pronounced shift may dominate.
    \item Several MPCs following an exponentially decaying power–delay profile, each with a distinct Doppler shift, yield to a selectivity that broadly follows the relative MPCs' powers.
    \item Multiple MPCs whose underlying rays overlap at the receiver, lead to a combined and irregular Doppler spectrum.
\end{enumerate}
While the first two cases are well documented, the last one introduces strong interference and uneven power distribution. In the frequency domain, this irregular Doppler spectrum leads to substantial leakage with a non-uniform power distribution, thereby complicating Doppler estimation and pilot interpolation. Here, the resolving power is largely determined by the symbol duration and pilot density.
%%%%%%%%%%%%%%%%%
 \begin{figure*}[t]
     \centering
\includegraphics[width=1\textwidth,height=6.2cm]{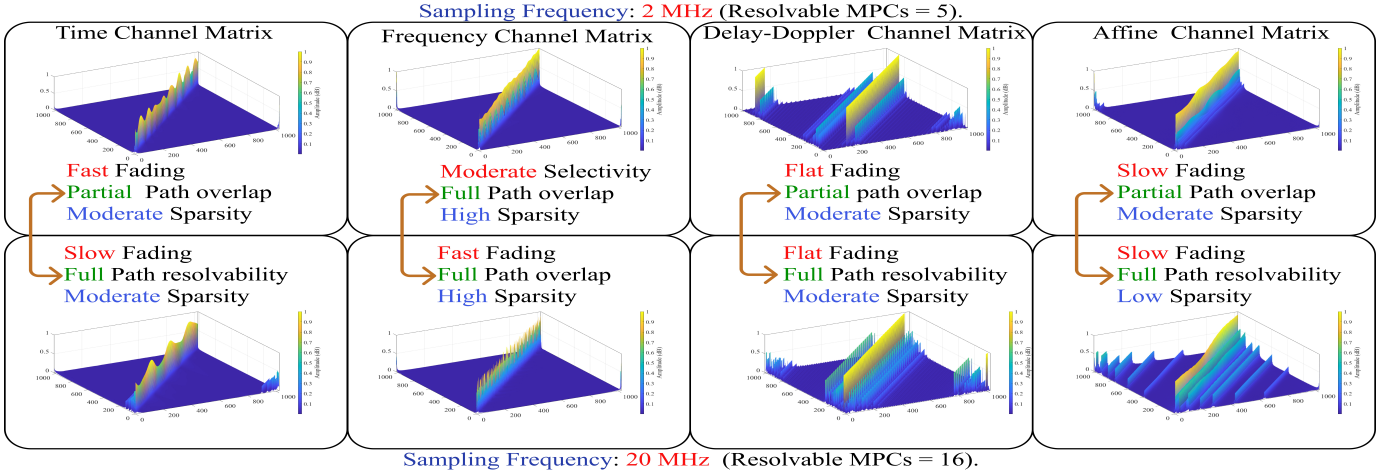}
\caption{Different domains' effective channel representation under 3GPP CDL channel model with Doppler Jake's spectrum.}
\label{fig: Channel_domains}
\end{figure*}
%%%%%%%%%%%%%%%%%%%%
%%%%%%%%%% subsection %%%%%%%%%%%%%%%%%%%%%%
\subsection{Emerging Domains}
Fast time-varying propagation environments fundamentally challenge the assumptions of stationarity and quasi-stationarity that underpin many classical channel models. High mobility induces intrinsic delay–Doppler coupling, with particularly pronounced effects under non-linear motion profiles such as \ac{UAV} trajectories. In parallel, the adoption of ultra-wide transmission bandwidths (exceeding $200$ MHz) significantly enhances delay resolution, thereby increasing the number of resolvable MPCs. Collectively, these trends modify the effective sparsity and spreading characteristics of the channel and undermine the dual interpretability of channel behavior in legacy time and frequency domains. This motivates the introduction of alternative logical domains that improve analytical tractability by promoting channel sparsity and \ac{MPCs} resolvability.

\par\textbf{Delay–Doppler Domain:}
Inspired by the channel spreading function, the \ac{DD}-domain provides a multiplexing and representation framework in which localized basis functions aim to yield sparse, weakly correlated, resolvable, and non-fading MPCs associated with distinct delay and Doppler indices. Under idealized conditions, this representation produces a full-rank, block-structured channel matrix that is well suited for both communication and sensing operations. The DD-domain attains high Doppler resolution by capturing small frequency shifts through extended temporal observation windows. However, the effectiveness of this representation relies on restrictive propagation assumptions, including ultra-wide bandwidths and a limited number of statistically independent scatterers with exponentially decaying power profile. These conditions are most closely satisfied in short-range mmWave scenarios and are less representative of dense or clustered propagation environments. In the presence of clustered arrivals, multiple physical rays may collapse into a single apparent MPC, thereby increasing inter-component correlation and inducing fractional Doppler leakage. Moreover, while long observation windows improve Doppler resolution for sensing applications, they also expose path birth–death dynamics that destabilize the apparent sparsity structure over time~\cite{ma2024impacts}. Delay dispersion in the DD domain further exhibits a strong dependence on system bandwidth. Narrowband signals result in fractional delay spreading and pulse leakage, whereas increasing bandwidth enhances MPC resolvability at the cost of a more distributed channel representation. Consequently, the interpretation of DD-domain channel responses depends critically on whether:
\begin{itemize}
    \item The majority of channel energy remains concentrated in a limited number of components.
    \item Mutual statistical dependence among these components is sufficiently weak in both delay and Doppler axes.
    \item The chosen pulse shaping and windowing functions have adequate separability of pulses \cite{bayat2025unified}.
    %\item The induced leakage in the DD domain is effectively controlled.
\end{itemize}
\par \textbf{Affine Domain:} Analogous to the DD domain, the affine domain aims to provide a sparse channel representation with resolvable, non-fading \ac{MPCs} while coupling the delay and Doppler into a scaled shift controlled by a design parameter \cite{11411756}. This scaling creates virtual channel segments whose span reflects maximum Doppler shift and whose partitioning reflects maximum delay. The MPCs spacing then follows the chosen scale, often enlarging the overall spread while preserving energy concentration. Notably, the affine domain possesses the unique capability of transforming an underspread channel into an overspread channel without compromising the channel sparsity level. However, the fractional components introduce an inter-segment interference and overlapping. This interference causes  strong fading between channel MPCs, requiring large guard intervals between each segment which comes at a significant reduction in spectral efficiency. Therefore, an effective use of the affine representation rests on four checks:
%%%%%%%%%%%%%% Figure %%%%%%%%%%%%%%%%%%%
\begin{itemize}
    \item Align the scaling with maximum dominant Doppler to keep energy focused within acceptable range.
    \item Design the guards size to cap the segment-edge leakage to a tolerable value.
    \item Choose segmentation that delivers only the Doppler–lag granularity the scene requires.
    \item Keep inter-segment correlation low so components remain independently resolvable.
\end{itemize}
%%%%%%%% Figure %%%%%%%%%%%%%%%% 
% \begin{figure}[t]
%     \centering
%     \includegraphics[scale=0.20]{figures/domains_sensing.jpg}
%     \caption{\textcolor{black}{Sensing performance of different multiplexing domains; (a) range profile, (b) velocity profile}}
%     \label{fig: sensing}
% \end{figure}

%  \begin{figure*}[t]
%      \centering
%   \includegraphics[scale=0.8]{figures/rr.jpg}
% \caption{}
% \label{fig: Channel_domains}
% \end{figure*}
\begin{figure*}[t]
    \centering
    \includegraphics[width=1\textwidth,height=6.5cm]{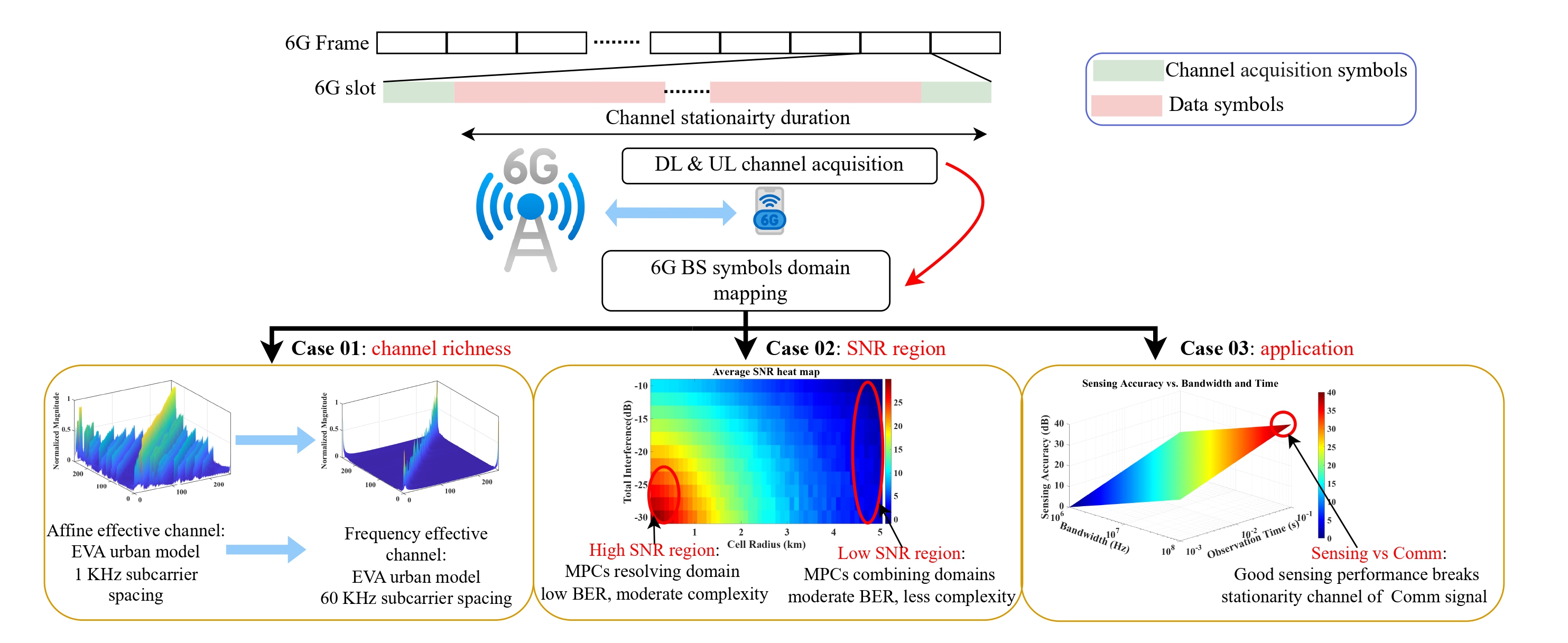}
    \caption{Interchanged-domain frame concept with adaptation drivers: channel richness, SNR region, and application constraints.}
    \label{fig:proposal}
\end{figure*}
%%%%%%%%%%%%%%%%%%%%
These logical domains provide joint delay and Doppler structural representation that time or frequency domains fail to achieve. However, the obtained DD information often has varying degrees of MPCs spread and ambiguous interpretation of the reflection sources, complicating the modeling of channel sparsity and MPC resolvability. For instance, due to the high Doppler resolution in the DD domain, fractional sampling of a single tap introduces a spreading behavior similar to the one originating from multiple integer Doppler shifts. However, effective MPCs are only resolvable and independent in the latter case, whereas the information within the fractional spread remains highly correlated. In terms of sparsity, the former type of spread can be mitigated using various fractional sampling suppression techniques, while the latter is inherently constrained by the maximum spread of the channel. Interpreting each representation through \ac{MPCs} energy distribution, mutual correlation, and leakage prevent sparse-looking effective channels from being mistaken for genuine resolvability. Fig.~\ref{fig: Channel_domains} illustrates the interchangeability of sparsity, resolvability, and \ac{MPCs} phase selectivity between effective channels of different domains and their dependability on the receiver sampling frequency as discussed in the previous section.
% \textcolor{black}{Figure~\ref{fig: Channel_domains} illustrates effective channel matrices across domains under a clustered 3GPP model: panels with correlated, overlapping MPCs and limited diagonal selectivity can look more “sparse” by energy concentration yet offer lower resolvability than panels that expose independent components.}

%%%%%%%%%%%%%% SECTION %%%%%%%%%%%%%%%%%%%%%%%
% \section{6G Interaction with Channel Resolvability and Separability}
% Although the threefold interaction between channel spread, sparsity, and MPC resolvability in the previously mentioned domains is complex and not well-defined, it offers a foundation for reshaping 6G application accordingly. This section interprets 6G's use-cases through the \ac{EVA} channel model as a running study-case, using the following pillars:
% \begin{itemize}
%     \item Energy concentration levels in the dominant \ac{MPCs}.
%     \item Mutual correlation between channel components.
%     \item Resolution dependent \ac{MPCs} resolvability (delay from bandwidth and Doppler from coherent frame time).
%     \item Processing-induced leakage such as fractional delay and Doppler, windowing tails, and segment edges.
% \end{itemize}

\section{6G Interaction with Channel Resolvability and Separability}
In this section, we propose a representative framework which maps the signal design and frame processing to different domains based on the desired channel behavior in these respective domains, and the associated applications, as illustrated in Fig. \ref{fig:proposal}
\subsection{Proposed Adaptive Interchanged-Domain Frame}
Channel information exchange between a gNB and a user-end (UE) provides a channel state view that can be exploited in more than one manner, depending on the propagation regime and the receiver operating conditions. A key limitation of committing to a single multiplexing and processing domain is that it implicitly assumes a fixed level of delay and Doppler resolvability, which is rarely maintained across mobility, bandwidth, and scattering conditions. In rich scattering, joint resolvability in delay and Doppler can be difficult to achieve. Typical examples include scenarios where Doppler structure is effectively compressed under narrowband operation, or where wider bandwidth reveals a large normalized delay spreading. In such scenarios, MPC resolution becomes ill-conditioned due to several practical limitations: spectral leakage and fractional spreading cause energy dispersion, inter-component coupling increases as MPCs become increasingly unresolvable within the system's delay resolution limit, and the apparent sparsity of the delay-domain channel representation is distorted, undermining the validity of sparse recovery frameworks. Therefore, the proposed framework focuses on these principles:
\begin{itemize}
    \item \textbf{Channel scattering richness governing domain selection}: In rich-scattering environments, MPCs exhibit pronounced mutual correlation, influence the shape of the Doppler power spectral density, and are particularly susceptible to fractional Doppler and delay shifts. These characteristics collectively elevate the complexity of MPC resolvability, as fractional spreading induces \ac{ISI} and \ac{ICI} that is inherently unstructured and resistant to closed-form modeling. The resulting detection ambiguity renders individual MPC resolution unreliable; consequently, it is often preferable to aggregate the contributing MPCs into an equivalent single-tap, unresolvable channel model. Under this formulation, the normalized delay and Doppler spread are explicitly constrained, allowing the aggregate channel effect to be absorbed into a compact representation while preserving reliable link performance, as in \ac{OFDM}. In contrast, when the channel response is dominated by a sparse set of well-separated MPCs characterized by negligible inter-component interference and low mutual correlation, the notion of resolvability becomes both meaningful and exploitable. In this regime, the receiver can coherently combine the resolved multipath contributions, yielding substantial diversity gains and a marked improvement in \ac{BER} performance through constructive multipath aggregation. Therefore, under this scenario, affine and \ac{DD} domain processing is preferred.
    \item \textbf{The preferred representation is \ac{SINR} region dependent}: hardware impairments, pulse leakage, UE cell position, and the noise variance reduce the effective SINR, thereby degrading the apparent sparsity and increasing inter-component correlation. Consequently, different \ac{SINR} regions naturally favor different choices. At low effective \ac{SINR}, domains where \ac{MPCs} cannot be resolved are preferred, while enhancing \ac{BER} performance through channel coding. Whereas at higher effective SINR, resolving MPCs becomes more advantageous as low interference allows relaxing complexity constrains in these domains while preserving reliable BER.
    \item \textbf{The degree of separation is application driven}: sensing typically requires a wide bandwidth and a large time observation window, which may introduce birth and death behavior and excessive delay spreading for communication UEs. Therefore, the suitable representation and the level of component separation should be selected according to the primary application.
\end{itemize}
% Fig. \ref{fig:proposal} illustrates different case scenarios on how domain mapping is interchanged based on channel behavior in different domains.  

\subsection{Use-Cases Implications Under EVA:}
Under EVA’s clustered structure and mid-band mobility, these behaviors alternate across different domains in ways that are relevant for communications, sensing, and security. By shifting from a single domain approach to an adaptive, interchanged-domain radio framework that responds to channel interactions across domains, new opportunities emerge for communication, sensing, and security. \\
\begin{figure*}[t]
    \centering
    \begin{subfigure}[b]{0.32\textwidth}
        \centering
        \includegraphics[width=\textwidth]{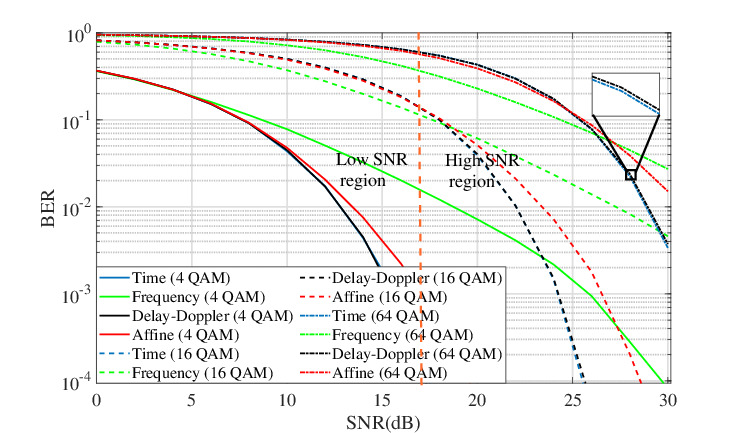}
        \caption{Communication performance of different multiplexing domains.}
        \label{fig:se_raw}
    \end{subfigure}
    \hfill
    \begin{subfigure}[b]{0.32\textwidth}
        \centering
        \includegraphics[width=\textwidth]{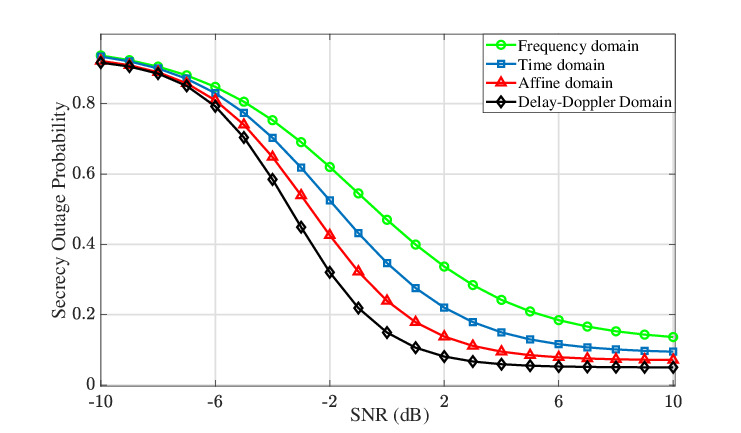}
        \caption{Secrecy outage probability from channel information in different domains.}
        \label{fig:se_eff}
    \end{subfigure}
        \begin{subfigure}[b]{0.32\textwidth}
        \centering
        \includegraphics[width=\textwidth]{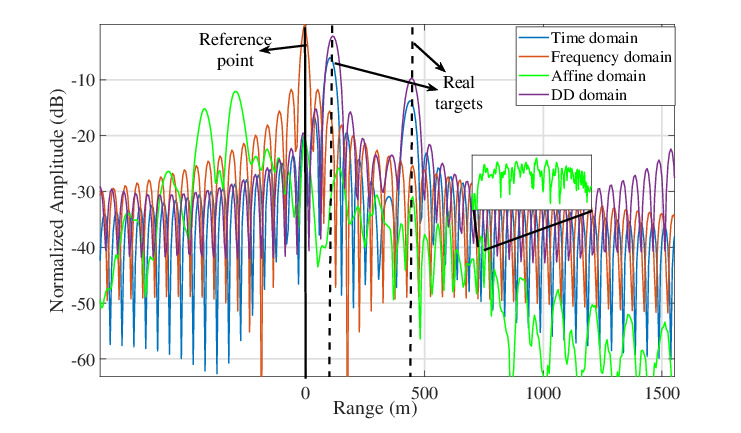}
        \caption{Sensing performance for different domains a range cut}
        \label{fig:range_cut}
    \end{subfigure}
    \caption{Evaluation performance of the 6G pillars showcasing the performance variety between different domains.}
    \label{fig:6G pillars}
\end{figure*}
\textbf{Communication}: Generally, channel equalization's complexity and reliability depend on the effective channel matrix's sparsity. In sparse conditions, such as static frequency-domain scenes with a near-diagonal structure, one tap equalization is sufficient, as seen in OFDM. Weakly correlated, resolvable components provide independent information and genuine diversity, enhancing the effective \ac{SNR} beyond classical spatial techniques. However, as sparsity decreases, the channel matrix transitions to multi-diagonal structures, shifting detection towards iterative deconvolution, diagonalization, and \ac{LS} and \ac{MMSE} techniques, leading to high computational complexity that rapidly escalates with matrix size. To reduce interference between data and pilots due to spreading, guard intervals are implemented in multiplexing domains, which degrade spectral efficiency in proportion to the level of dispersion. An ideal next-generation generalized receiver should aim to extract channel information in areas of high concentration and low correlation, applying equalization where instantaneous SNR is optimally conditioned, and placing pilots where observed spread is minimal. Under \ac{EVA} mid-band mobility, frequency-domain interpolation effectively recovers the response for static or low mobility, while data can be multiplexed in time, affine, or delay–Doppler domains where \ac{MPCs} remain resolvable. Coherent combining of these components offers higher-order diversity. Moreover, taps resolving domains leverage their inherent diversity to enhance \ac{BER} compared to frequency domain approaches. Nonetheless, as modulation order increases, diversity gains diminish at low SNR values, causing convergence across all domains as seen in Fig. \ref{fig:6G pillars} (a).\\
\textbf{Security}: \ac{PLS} emphasizes on three main channel traits: diversity, uniqueness, and richness, which influence its resilience against interference, authentication, and key generation entropy, respectively. It follows that selecting the appropriate domain where the channel is observed can enhance these qualities. Specifically, domains that spread data symbols across time and frequency resources (such as affine and DD) demonstrate greater resistance to narrow-band jamming, preserving recoverable information and enhancing effective \ac{SNR} through diversity combining. In contrast, frequency multiplexing is vulnerable to narrow-band interference, risking total information loss unless mitigated by redundancy or interleaving, which reduces spectral efficiency. Channel characteristics vary across domains; for instance, in the frequency domain, the overlapping channel coefficients limit the degree of freedom (DoF). Conversely, DD and affine domains provide richer identification capabilities owing to their delay and Doppler tuples and correlation structures. The affine domain additionally benefits from an extra DoF through varied phase shifts, but is also more prone to channel correlation in environments with rich scattering, such as urban \ac{EVA} models, which diminishes independent information and entropy for key generation. This relationship is visually represented in Fig. \ref{fig:6G pillars} (b), illustrating the secrecy outage probability across different domains in a time-varying channel. It concludes that an increase in channel parameter resolvability correlates with enhanced information entropy, with the DD domain yielding the most favorable secrecy outage probability.\\
% \subsubsection{Sensing}
%  \begin{figure}[t]
%     \centering
%     \includegraphics[width=\columnwidth,height=0.45\textheight,keepaspectratio]{figures/SOP.jpg}
%     \caption{Secrecy outage probability from channel information in different domains.}
%     \label{fig: security}
% \end{figure}
% \begin{figure}[t]
%     \centering
%     \begin{subfigure}[b]{0.5\textwidth}
%         \centering
%         \includegraphics[width=\textwidth]{figures/new_range_cut.jpg}
%         \caption{}
%         \label{fig:se_raw}
%     \end{subfigure}
%     \hfill
%     \begin{subfigure}[b]{0.5\textwidth}
%         \centering
%         \includegraphics[width=\textwidth]{figures/new_Doppler_cut.jpg}
%         \caption{}
%         \label{fig:se_eff}
%     \end{subfigure}
%     \caption{Sensing performance for different domains: (a) Range cut, (b) Doppler cut.}
%     \label{fig:Sensing}
% \end{figure}
\textbf{Sensing}: Although \ac{ISAC} seeks to unify sensing and communication, they interpret the channel reflections differently. Specifically, the interplay between sparsity, resolvability, and spread differs in sensing channels, which are typically sparse due to directional antennas and \ac{RCS}-based filtering. However, isolating only the legitimate target is challenging. Even under \ac{LoS}, \ac{MPCs} can generate ghost targets and raise false alarms, particularly in dense urban clutter where high-\ac{RCS} paths dominate~\cite{zhang2021enabling}. Alternatively, resolvability, which refers to resolution from a sensing perspective, defines the system's ability to distinguish between multiple targets. Therefore, resolvability depends on available resources: bandwidth sets range granularity, whereas coherent symbol duration sets velocity granularity. Nevertheless, closely correlated \ac{MPCs} from the same cluster may still produce ghosts, which is especially critical in \ac{V2X}. Furthermore, delay spreading is governed by the round-trip distance to the farthest detectable target; a larger spread requires longer pulse repetition intervals and, as a result, introduces phase ambiguity that can fold higher Doppler shifts into lower ones. Given that sensing merges time and frequency processing into a two-dimensional framework, other domains that spread in time and frequency resemble sensing channels under optimal sparsity and resolvability. For instance, the \ac{DD} domain directly corresponds to the range-Doppler map, whereas the affine domain encapsulates coupled sensing effects, which can be decoupled through transform design~\cite{10891132}. Beyond their inherent sensing capabilities, these domains offer strong performance in complex sensing scenarios. The \ac{DD} domain, with its high Doppler resolution, is well-suited for applications where multiple targets move in close proximity at similar speeds, enabling efficient multi-target tracking, particularly in \ac{V2X} applications. Fig.~\ref{fig:6G pillars} (c) sheds light on how the different domains provide sensing information for the range case, \textcolor{black}{where \ac{DD} and time domains provide a identical range information accuracy, whereas frequency domain is agnostic to range information. In addition, affine domain is affected by design parameters leading to ghost targets and pulse distortion.}

%%%%%%%%%%%%%%%%%%%%%%% OPEN ISSUES %%%%%%%%%%%%%%%%%%%%%%%%%%%%%%%%%%%%%%%
\section{Open Issues and Future directions}
% This section introduces the open issues, challenges and research directions for the resolvability, sparsity and spreading domain-based processing. 
\subsection{ THz and mmWave Frequencies}
High carrier frequencies offer wider bandwidths and narrower beams, which improve multipath resolvability but reduce channel coherence time and intensify leakage arising from fractional Doppler, phase noise, and beam squint \cite{tera_hertz}. Moreover, wider bandwidths increase MPCs resolvability, which in turn enlarges pilot and guard overhead requirements. A practical approach is to design guards to a target leakage level, partition the wideband spectrum into subbands, and combine short, robust frames for communication with longer observation frames for sensing.

\subsection{Hardware Impairments}
Phase noise, carrier frequency offset, and in-phase/quadrature imbalance degrade sparsity and resolvability across different domains, while power amplifier nonlinearity and quantization noise elevate leakage floors. These impairments introduce ambiguity in modeling MPCs behavior within each representation. In particular, the \ac{DD} domain is highly sensitive to fractional timing-offset induced leakage, even when MPCs align with integer bins, a phenomenon that is largely mitigated in the frequency domain. Consequently, performance evaluations should consider these limitations, rather than restricting analysis to idealized conditions.

\subsection{AI Receivers: Training Under Spread and Resolvability}
Data-driven receivers are inherently sensitive to variations in channel representations. A channel that exhibits rapid fluctuations in the frequency domain and requires frequent online adaptation may appear more stationary in the \ac{DD} domain, thereby reducing the need for continuous fine-tuning \cite{AI_update}. Moreover, effective channel sparsity enables learning-based receivers to exploit richer channel variations, which is advantageous for applications such as high-sensitivity target sensing.

%%%%%%%%%%%%%%%%%%%%%%%%% Conclusion %%%%%%%%%%%%%%%%%%%%%%%%%%%%
\section{Conclusion}
The shift toward 6G necessitates a departure from treating channel sparsity and resolvability as static physical constants. While legacy time-frequency frameworks struggle under the non-stationary conditions of high-mobility scenarios, \ac{DD} and affine representations can offer improved resolution and diversity, provided that processing induced leakage and inter-component correlation are managed. Across the communication, sensing, and security pillars, no single multiplexing domain is universally optimal. Instead, the preferred representation depends on the channel richness, SINR, and application: the DD domain supports high-resolution velocity tracking for \ac{ISAC}, while the affine domain offers additional degrees of freedom for \ac{PLS} and robustness under non-linear motion. Ultimately, 6G air interfaces will benefit from domain agility. An adaptive interchanged domain radio architecture can move beyond a "one-fits-all" waveform by projecting processing into the domain where the channel is most favorable under the available resolvability and leakage constraints, supporting ultra-reliable communication and high-granularity sensing.
% \section{Conclusion}
% The shift toward 6G necessitates a departure from treating channel sparsity and resolvability as static physical constants. While legacy time-frequency frameworks struggle under the non-stationary conditions of high-mobility 6G scenarios, emerging delay-Doppler and affine representations offer superior resolution and diversity, given that processing-induced leakage and inter-component correlation are managed. Our investigation across communication, sensing, and security pillars reveals that no single multiplexing domain is universally optimal. Instead, the "best" representation fluctuates with the environment: the delay-Doppler domain excels in high-resolution velocity tracking for ISAC, while the affine domain provides unique degrees of freedom for physical layer security and robustness against non-linear motion. Ultimately, the future of 6G air interfaces lies in domain-agility. By moving toward an adaptive interchanged-domain radio architecture, we can move beyond the constraints of a "one-fits-all" waveform. Such a framework, guided by the four pillars of energy concentration, correlation, resolving power, and leakage, allows the network to dynamically project signal processing into the domain where the channel is most sparse and the information is most secure. This evolution will be fundamental in achieving the ultra-reliable, high-granularity sensing and communication goals of the next decade.
%%%%%%%%%%%%%%%%%%%%%%%%%%%%%%%%%%%%%%%%%%%%%%%%%%%%%%%%%%%%%%%%%
\bibliographystyle{IEEEtran}
\bibliography{Mendeley}
\end{document}

%% file: acro.tex
\DeclareAcronym{RIS}{
  short = RIS ,
  long  = reconfigurable intelligent surfaces ,
}
\DeclareAcronym{6G}{
  short = 6G,
  long  = sixth generation ,
}
\DeclareAcronym{5G}{
  short = 5G,
  long  = 5th generation ,
}
\DeclareAcronym{RDM}{
  short =   RDM,
  long  = range-Doppler map ,
}
\DeclareAcronym{ISAC}{
  short = ISAC ,
  long  = integrated sensing and communication,
}
\DeclareAcronym{SA-OFDM}{
  short = SA-OFDM ,
  long  = subcarrier aliasing OFDM,
}
\DeclareAcronym{EM}{
  short = EM ,
  long  = electromagnetic,
}
\DeclareAcronym{BS}{
  short = BS,
  long  = base station ,
}
\DeclareAcronym{UE}{
  short = UE ,
  long  = user equipment ,
}
\DeclareAcronym{MISO}{
  short = MISO,
  long  = multiple-input single-output ,
}
\DeclareAcronym{MMSE}{
  short = MMSE ,
  long  = minimum mean squared error ,
}
\DeclareAcronym{RMSE}{
  short = RMSE ,
  long  = root mean squared error ,
}
\DeclareAcronym{DFT}{
  short = DFT,
  long  = discrete Fourier transform ,
}
\DeclareAcronym{IDFT}{
  short = IDFT,
  long  = inverse discrete Fourier transform ,
}
\DeclareAcronym{FFT}{
  short = FFT,
  long  = fast Fourier transform ,
}
\DeclareAcronym{ISFFT}{
  short = ISFFT,
  long  = inverse symplectic finite Fourier transform ,
}
\DeclareAcronym{IFFT}{
  short = IFFT,
  long  = inverse fast Fourier transform ,
}
\DeclareAcronym{SFFT}{
  short = SFFT,
  long  = symplectic finite Fourier transform ,
}
\DeclareAcronym{THz}{
  short = THz,
  long  = Terahertz ,
}
\DeclareAcronym{IoT}{
  short = IoT,
  long  = internet of things ,
}
\DeclareAcronym{V2X}{
  short = V2X,
  long  = vehicle to everything ,
}
\DeclareAcronym{V2V}{
  short = V2V,
  long  = vehicle to vehicle ,
}
\DeclareAcronym{MSE}{
  short = MSE,
  long  = mean square error ,
}
\DeclareAcronym{CSI}{
  short = CSI ,
  long  = channel state information ,
}
\DeclareAcronym{MIMO}{
  short = MIMO,
  long  = multiple-input multiple-output ,
}
\DeclareAcronym{UPA}{
  short = UPA,
  long  = uniform planner array ,
}
\DeclareAcronym{RF}{
  short = RF,
  long  = radio-frequency ,
}
\DeclareAcronym{mmWave}{
  short = mmWave,
  long  = millimeter-wave ,
}
\DeclareAcronym{AoA}{
  short = AoA ,
  long  = angle of arrival ,
}
\DeclareAcronym{AoD}{
  short = AoD,
  long  = angle of departure ,
}
\DeclareAcronym{EKF}{
  short = EKF,
  long  = extended Kalman filter ,
}
\DeclareAcronym{LMS}{
  short = LMS,
  long  = least mean square ,
}
\DeclareAcronym{BiLMS}{
  short = BiLMS,
  long  = bi-directional LMS ,
}
\DeclareAcronym{SNR}{
  short = SNR,
  long  = signal-to-noise ratio ,
}
\DeclareAcronym{LoS}{
  short = LoS,
  long  = line-of-sight ,
}
\DeclareAcronym{NLoS}{
  short = NLoS,
  long  = non line-of-sight ,
}
\DeclareAcronym{TDD}{
  short = TDD,
  long  = time-division duplexing ,
}
\DeclareAcronym{NMSE}{
  short = NMSE,
  long  = normalized mean square error ,
}
\DeclareAcronym{PAPR}{
  short = PAPR,
  long  = peak to average power ratio ,
}
\DeclareAcronym{ISI}{
  short = ISI,
  long  = inter-symbol interference ,
}
\DeclareAcronym{ICI}{
  short = ICI,
  long  = inter-carrier interference ,
}
\DeclareAcronym{SDR}{
  short = SDR,
  long  = semidefinite relaxation ,
}
\DeclareAcronym{QoS}{
  short = QoS,
  long  = quality of service ,
}
\DeclareAcronym{NOMA}{
  short = NOMA,
  long  = non-orthogonal multiple access ,
}
\DeclareAcronym{OMA}{
  short = OMA,
  long  = orthogonal multiple access ,
}
\DeclareAcronym{NU}{
  short = NU,
  long  = near user ,
}
\DeclareAcronym{CIR}{
  short = CIR,
  long  = channel impulse response ,
}
\DeclareAcronym{FU}{
  short = FU,
  long  = far user ,
}
\DeclareAcronym{CP}{
  short = CP,
  long  = cyclic prefix ,
}
\DeclareAcronym{ZP}{
  short = ZP,
  long  = zero padding ,
}
\DeclareAcronym{ZS}{
  short = ZS,
  long  = zero suffix ,
}
\DeclareAcronym{ZF}{
  short = ZF,
  long  = zero forcing ,
}
\DeclareAcronym{RCP}{
  short = RCP,
  long  = reduced-CP ,
}
\DeclareAcronym{FZS}{
  short = FZS,
  long  = full-ZS ,
}
\DeclareAcronym{RZP}{
  short = RZP,
  long  = reduced-zero padded ,
}
\DeclareAcronym{FCP}{
  short = FCP,
  long  = full-CP ,
}
\DeclareAcronym{RFCP}{
  short = RFCP,
  long  = reduced-full-CP ,
}
\DeclareAcronym{BER}{
  short = BER,
  long  = bit error rate ,
}
\DeclareAcronym{SIC}{
  short = SIC,
  long  = successive interference cancellation ,
}
\DeclareAcronym{PLS}{
  short = PLS,
  long  = physical layer security ,
}
\DeclareAcronym{MRT}{
  short = MRT,
  long  = maximum ratio transmission ,
}
\DeclareAcronym{AWGN}{
  short = AWGN,
  long  = additive white Gaussian noise,
}
\DeclareAcronym{SINR}{
  short = SINR,
  long  = signal-to-interference-plus-noise ratio ,
}
\DeclareAcronym{BPSK}{
  short = BPSK,
  long  = binary phase shift keying ,
}
\DeclareAcronym{QPSK}{
  short = QPSK,
  long  = quadrature phase shift keying ,
}
\DeclareAcronym{SVD}{
  short = SVD,
  long  = singular value decomposition ,
}
\DeclareAcronym{EVD}{
  short = EVD,
  long  = eigenvalue decomposition ,
}
\DeclareAcronym{PDF}{
  short = PDF,
  long  = probability density function ,
}
\DeclareAcronym{SER}{
  short = SER,
  long  = symbol error rate ,
}
\DeclareAcronym{MGF}{
  short = MGF,
  long  = moment generating function ,
}
\DeclareAcronym{2D}{
  short = 2D,
  long  = two-dimensional ,
}
\DeclareAcronym{3D}{
  short = 3D,
  long  = three-dimensional ,
}
\DeclareAcronym{CLT}{
  short = CLT,
  long  = central limit theorem ,
}
\DeclareAcronym{QAM}{
  short = QAM,
  long  = quadrature amplitude modulation ,
}
\DeclareAcronym{SISO}{
  short = SISO,
  long  = single-input single-output ,
}
\DeclareAcronym{CE}{
  short = CE,
  long  = channel estimation ,
}
\DeclareAcronym{KG}{
  short = $K_G$,
  long  = generalized-K ,
}
\DeclareAcronym{LSKRF}{
  short = LSKRF,
  long  = least squares Khatri-Rao factorization ,
}
\DeclareAcronym{FMCW}{
  short = FMCW,
  long  = frequency modulated continuous wave ,
}
\DeclareAcronym{FSK}{
  short = FSK,
  long  = frequency shift keying ,
}
\DeclareAcronym{JSAC}{
  short = JSAC,
  long  = joint sensing and communication ,
}
\DeclareAcronym{OTFS}{
  short = OTFS,
  long  = orthogonal time-frequency space ,
}
\DeclareAcronym{OTSM}{
  short = OTSM,
  long  = orthogonal time sequency multiplexing ,
}
\DeclareAcronym{ADAS}{
  short = ADAS,
  long  = Advanced driver assistant system ,
}
\DeclareAcronym{MP}{
  short = MP,
  long  = message passing ,
}
\DeclareAcronym{ML}{
  short = ML,
  long  = maximum likelihood ,
}
\DeclareAcronym{OFDM}{
  short = OFDM,
  long  = Orthogonal frequency division multiplexing ,
}
\DeclareAcronym{PSD}{
  short = PSD,
  long  = power spectral density,
}
\DeclareAcronym{AFDM}{
  short = AFDM,
  long  = Affine frequency division multiplexing ,
}
\DeclareAcronym{OCDM}{
  short = OCDM,
  long  = orthogonal chirp division multiplexing ,
}
\DeclareAcronym{ODDM}{
  short = ODDM,
  long  = orthogonal delay-Doppler division multiplexing,
}
\DeclareAcronym{LTV}{
  short = LTV,
  long  = linear time-varying,
}
\DeclareAcronym{LTI}{
  short = LTI,
  long  = linear time-invarying,
}
\DeclareAcronym{NTN}{
  short = NTN,
  long  = non-terrestrial network,
}

\DeclareAcronym{AI}{
  short = AI,
  long  = artificial intelligence,
}
\DeclareAcronym{LEO}{
  short = LEO,
  long  = low Earth orbit,
}
\DeclareAcronym{TF}{
  short = T-F,
  long  = time-frequency,
}
\DeclareAcronym{DD}{
  short = DD,
  long  = delay-Doppler,
}
\DeclareAcronym{AF}{
  short = AF,
  long  = ambiguity function,
}
\DeclareAcronym{LoRa}{
  short = LoRa,
  long  = Long-Range,
}
\DeclareAcronym{EVA}{
  short = EVA,
  long  = extended vehicular A,
}
\DeclareAcronym{SMT}{
  short = SMT ,
  long  = sparse multi-tap,
}
\DeclareAcronym{TDL}{
  short = TDL ,
  long  = tapped delay line,
}
\DeclareAcronym{UAV}{
  short = UAV ,
  long  = unmanned aerial vehicular,
}
\DeclareAcronym{MPCs}{
  short = MPCs ,
  long  = multipath components,
}

\DeclareAcronym{LS}{
  short = LS,
  long  = least squares,
}
\DeclareAcronym{KPIs}{
  short = KPIs,
  long  = key performance indicators,
}
\DeclareAcronym{RCS}{
  short = RCS,
  long  = radar cross section,
}
\DeclareAcronym{MRC}{
  short = MRC,
  long  = maximum ratio combiner,
}